\documentclass[useAMS,usenatbib,usegraphicx]{mn2e}

\usepackage{color}
\usepackage{amssymb}
\usepackage{times}

\newcommand\aj{{AJ}}
\newcommand\apj{{ApJ}}
\newcommand\apjl{{ApJ}}
\newcommand\apss{{Ap\&SS}}
\newcommand\aap{{A\&A}}
\newcommand\mnras{{MNRAS}}
\newcommand\pasp{{PASP}}
\newcommand\pasj{{PASJ}}

\title[New AM CVn binaries from SDSS-III]{Two new AM Canum Venaticorum binaries from the Sloan Digital Sky Survey III}
\author[P. J. Carter et al.]{P. J. Carter,$^{1}$\thanks{E-mail: philip.carter@warwick.ac.uk} B. T. G\"{a}nsicke,$^{1}$ D. Steeghs,$^{1}$ T. R. Marsh,$^{1}$ E. Breedt,$^{1}$ T. Kupfer,$^{2}$ \newauthor N. P. Gentile Fusillo,$^{1}$ P. J. Groot$^{2}$ and G. Nelemans$^{2,3}$\\
$^{1}$Department of Physics, University of Warwick, Coventry CV4 7AL, UK\\
$^{2}$Department of Astrophysics/IMAPP, Radboud University Nijmegen, PO Box 9010, 6500 GL Nijmegen, the Netherlands\\
$^{3}$Institute for Astronomy, KU Leuven, Celestijnenlaan 200D, 3001 Leuven, Belgium\\
}

\begin{document}

\date{Accepted/Received.}

\pagerange{\pageref{firstpage}--\pageref{lastpage}} \pubyear{2013}

\maketitle

\label{firstpage}

\begin{abstract}
The AM Canum Venaticorum (AM CVn) binaries are a rare group of ultra-short period, mass-transferring white dwarf binaries, some of which may be Type Ia supernova progenitors. More than a third of the total known population of AM CVn binaries have been discovered via the Sloan Digital Sky Survey (SDSS). Here, we discuss our search for new AM CVns in the SDSS spectroscopic data base, and present two new AM CVns discovered in SDSS-III spectroscopy\textcolor{black}{, SDSS\,J113732.32+405458.3 and SDSS\,J150551.58+065948.7.}
\textcolor{black}{The AM CVn binaries exhibit a connection between their spectral appearance and their orbital period, the spectra of these two new AM CVns suggest that they may be long period systems.}
Using the radial velocity variations of the emission lines, we measure a possible orbital period of 59.6\,$\pm$2.7\,minutes for SDSS\,J113732.32+405458.3.
Since our search of SDSS spectroscopy has revealed only these two new systems, it is unlikely that a large population of AM CVn binaries have been missed, and their discovery should have little effect on previous calculations of the AM CVn space density.
\end{abstract}



\section{Introduction}

The AM Canum Venaticorum (AM CVn) binaries are hydrogen-deficient, ultracompact, mass-transferring white dwarf binaries. This rare class of objects is characterised by their short orbital periods, which fall in the range of 5 to 65\,minutes, and their helium-rich spectra. The white dwarf primary accretes from a degenerate or semi-degenerate companion, allowing their periods to fall significantly below the expected period minimum of normal hydrogen-rich cataclysmic variables \citep[$\sim$80\,min;][]{1982ApJ...254..616R,2009MNRAS.397.2170G}. See \citet{2010PASP..122.1133S} for a recent review of the AM CVn binaries.

The AM CVn class is of particular interest due to their potential as progenitors of Type Ia supernovae \citep[e.g.][]{2009ApJ...699.1365S,2013arXiv1305.6925S}, and sub-luminous SN Ia-like explosions (`SN.Ia'; \citealt{2007ApJ...662L..95B,2011MNRAS.411L..31B}). \textcolor{black}{They are also extremely important due to their expected gravitational wave emission; AM CVn binaries are predicted to be some of the strongest known sources of low-frequency gravitational waves that would be detected by future space-borne gravitational wave detectors \citep{2006CQGra..23S.809S,2006MNRAS.371.1231R,2009CQGra..26i4030N,2012ApJ...758..131N}.}

\textcolor{black}{The gravitational wave emission thought to drive the evolution of the AM CVn binaries leads to a strong dependence of the mass accretion rate, and hence their observed properties, on the orbital period. The observational characteristics of an AM CVn binary are commonly separated into four distinct categories, based on a combination of empirical evidence and theoretical models. Direct impact accretion is expected for orbital periods below $\sim$10\,minutes (e.g.\ HM Cnc, \citealt{2010ApJ...711L.138R}; V407 Vul, \citealt{2002MNRAS.331L...7M}); in these systems no accretion disc forms.
Analogous to cataclysmic variables, where a white dwarf accretes from an H-rich disc, the He-rich discs found in AM CVn stars are subject to a thermal instability \citep{1981A&A...104L..10M,1997PASJ...49...75T,2012A&A...544A..13K,2012MNRAS.419.2836R}. At short orbital periods ($\sim$10 to $\sim$20\,min), the mass transfer rate is sufficiently high to keep the disc in a stable hot, bright state (e.g.\ HP Lib, \citealt{1994MNRAS.271..910O}; SDSS\,J1908+3940, \citealt{2011ApJ...726...92F}). At intermediate orbital periods, the discs undergo quasi-periodic outbursts, cycling between a quiescent faint state with a cool disc, observationally characterised by strong emission lines, and a hot bright state, characterised by absorption lines from the optically thick disc (e.g.\ CR Boo, \citealt{1987ApJ...313..757W}; SDSS\,J0129+3842, \citealt{2011arXiv1104.0107S}; PTF1\,J0943+1029, \citealt{2013MNRAS.430..996L}). At the longest orbital periods ($\gtrsim$45\,min), corresponding to the lowest mass transfer rates, 
the disc is thought to be in a stable faint state, and the spectra of these systems are hence dominated by strong emission lines (e.g.\ GP Com, \citealt{1981ApJ...244..269N}; SDSS\,J1552+3201, \citealt{2007MNRAS.382.1643R}).}

The first AM CVn binaries were discovered serendipitously in a variety of different ways \citep[e.g.][]{1971ApJ...170L..39B,1999A&A...349L...1I}, producing a small and heterogeneous sample. A total of six new AM CVn binaries were discovered in the Sloan Digital Sky Survey (SDSS; \citealt{2000AJ....120.1579Y}) I and SDSS-II spectroscopic data, via their helium emission dominated spectra \citep{2005MNRAS.361..487R,2005AJ....130.2230A,2008AJ....135.2108A}. These provided the first well understood, homogeneous sample, that allowed \citet{2007MNRAS.382..685R} to estimate the local space density by calibrating the theoretical population models. \citet{2007MNRAS.382..685R} obtained a value of 1--3~$\times$~10$^{-6}$ pc$^{-3}$. This was an order of magnitude lower than the expected space density at the time, derived from population synthesis (2~$\times$~10$^{-5}$ pc$^{-3}$, based on \citealt{2001A&A...368..939N}).

Since the known AM CVn binaries have been found to occupy a relatively sparsely-populated region of colour space, which has a low spectroscopic completeness in the SDSS data base, we have conducted a dedicated spectroscopic survey designed to uncover the `hidden' population of AM CVn binaries in the SDSS photometry. Details of this programme, including the sample selection criteria, and the seven AM CVns discovered, were presented by \citet{2009MNRAS.394..367R}, \citet{2010ApJ...708..456R} and \citet{2013MNRAS.429.2143C}.

We recently presented the latest results from this survey; where we discussed increasing our AM CVn detection efficiency using new colour--colour cuts based on photometry from \textit{GALEX}, and gave our revised estimate of the AM CVn space density, (5~$\pm$~3)~$\times$~10$^{-7}$~pc$^{-3}$ \citep{2013MNRAS.429.2143C}.

More than a third of the total known population of AM CVn binaries have been discovered via the SDSS, though recently they have also been found increasingly via synoptic surveys\textcolor{black}{, with the Catalina Real Time Transient Survey (CRTS; \citealt{2009ApJ...696..870D}), and the Palomar Transient Factory \citep[PTF;][]{2009PASP..121.1395L} significantly increasing the population 
\citep{2011ApJ...739...68L,2013ATel.4726....1W,2013MNRAS.430..996L}.}
Here we present two new systems discovered amongst the SDSS-III spectroscopic data. Motivated by their discovery, we have applied our proposed colour-cuts based on \textit{GALEX}, whilst extending the original colour-selection in the SDSS bands, to search for any further AM CVns in the most recent SDSS spectroscopy.


\section{A search of the SDSS spectroscopic data base}

\begin{figure*}
 \includegraphics[width=0.98\textwidth]{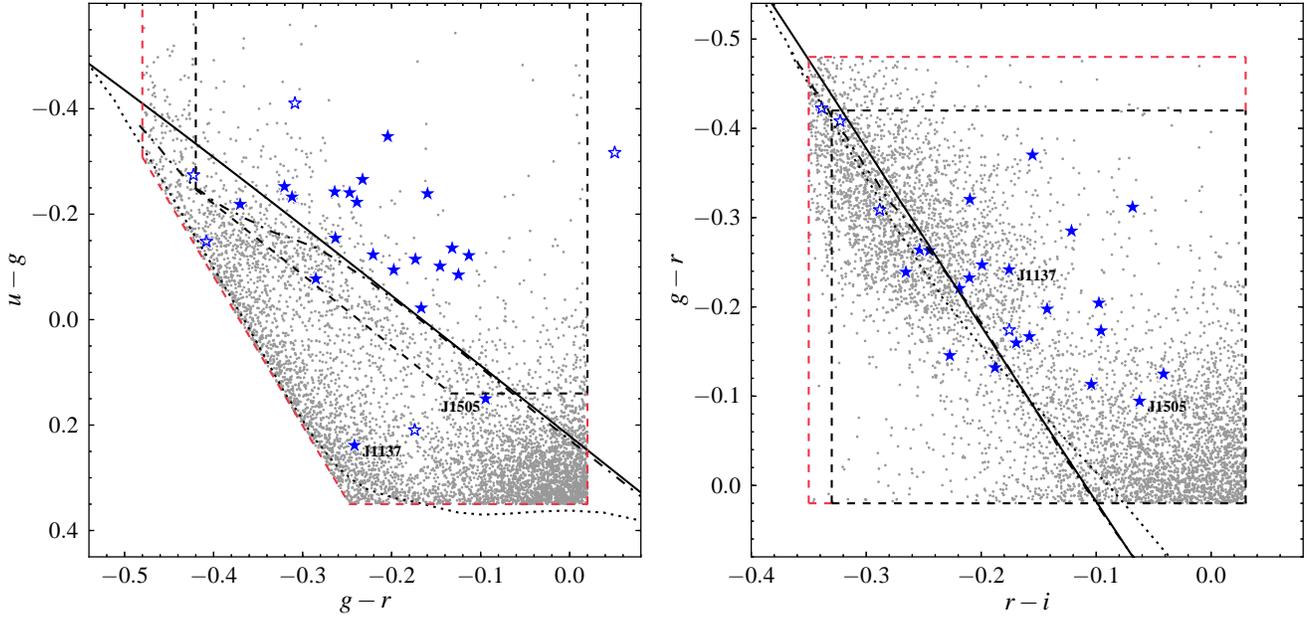}
 \caption{Dereddened colours of the known AM CVn binaries with $ugri$ photometry (blue stars). Those systems that have short periods, or photometry that may not correspond to their quiescent state are marked with open stars. The two new AM CVns are labelled. \textcolor{black}{Colours of the faint PTF-discovered systems are taken from \citet{2013MNRAS.430..996L}}. The black dashed lines show the \citet{2009MNRAS.394..367R} colour-cuts, the red dashed lines represent the extended cuts we use here. The objects selected are plotted as grey dots. The two AM CVns with $u-g$ outside the original cuts are the new systems discussed in this article. The solid line shows the blackbody cooling track, the dotted and dot-dashed lines represent the DA and DB white dwarf cooling tracks (log $g = 8.0$).\label{f:newcuts}}
\end{figure*}
The SDSS-III comprises four spectroscopic surveys, largely targeting quasars and galaxies \citep{2011AJ....142...72E}. Recent data releases have significantly increased the number of spectra available. We discovered the two new systems discussed in this article whilst examining spectra of objects targeted as white dwarfs by the SDSS-III. This motivated us to conduct a fuller search of the SDSS spectroscopic data base, examining objects observed as part of the SDSS-III with colours similar to the known AM CVns (the spectroscopic observations from the SDSS and SDSS-II have been well explored by previous studies, \citealt{2005MNRAS.361..487R,2005AJ....130.2230A,2008AJ....135.2108A}). 

We have previously exploited the similar colours of the known AM CVns to conduct a dedicated survey of objects with only photometry in the SDSS data base, to search for the `hidden' population \citep{2013MNRAS.429.2143C}. The two new AM CVns both have $u-g$ colours slightly outside the range occupied by the previously identified SDSS AM CVns, and so we have expanded \citet{2009MNRAS.394..367R}'s original colour selection for our new search of the spectroscopic data base, as shown in Fig. \ref{f:newcuts} and Table \ref{t:cuts}. We also apply our colour-cuts based on \textit{GALEX} NUV photometry, to increase the efficiency of our search, as these remove many of the subdwarfs and quasars from the sample \citep{2013MNRAS.429.2143C}. As described in \citet{2013MNRAS.429.2143C}, we retain objects that have extreme $\rmn{NUV}-u$ colours (as they may be variable), and those that do not have a match in the \textit{GALEX} data base. Our constraints are summarised in Table \ref{t:cuts}. We correct colours for Galactic extinction as described by \citet{2009MNRAS.394..367R} and \citet{2013MNRAS.429.2143C}.
\begin{table}
\centering
\caption{Colour selection used to search for AM CVns in the SDSS spectroscopic data base. We select only objects classified as a point source, with spectra that were made available in SDSS Data Releases 8, 9 or 10.}
\label{t:cuts}
\begin{tabular}{l l}
\hline
Colour		& Constraint \\
\hline
$u-g$		& $< 0.35$ \\
$u-g$		& $< 2.83\,(g-r) +1.05$ \\
$g-r$		& $< 0.02$ \\
$g-r$		& $> -0.48$ \\
$r-i$		& $< 0.03$ \\
$r-i$		& $> -0.35$ \\
$\rmn{NUV}-u$	& $> 4.34\,(g-r) + 0.5$ \ OR \ $< -1$\\
$\rmn{NUV}-u$	& $< 6.76\,(r-i) + 1.85$ \ OR \ $>1.5$\\
\hline
\end{tabular}
\end{table}

Applying these constraints to SDSS Data Release 10 (DR10), cross-matched with \textit{GALEX} Release 7, resulted in the selection of 5728 objects, for which the spectra were then visually inspected. The main contaminants in this colour-selection are hydrogen atmosphere (DA) white dwarfs, and quasars. We recovered five AM CVns with spectra taken as part of the SDSS-III surveys\footnote{The previously known AM CVns with new spectra in the SDSS-III are SDSS\,J0129+3842, SDSS\,J0804+1616 and SDSS\,J1721+2733.}, including the two new systems, but found no additional new AM CVns.


\section{The new AM CVn binaries}

\subsection{Average spectra}

%
\begin{figure*}
 \includegraphics[width=0.98\textwidth]{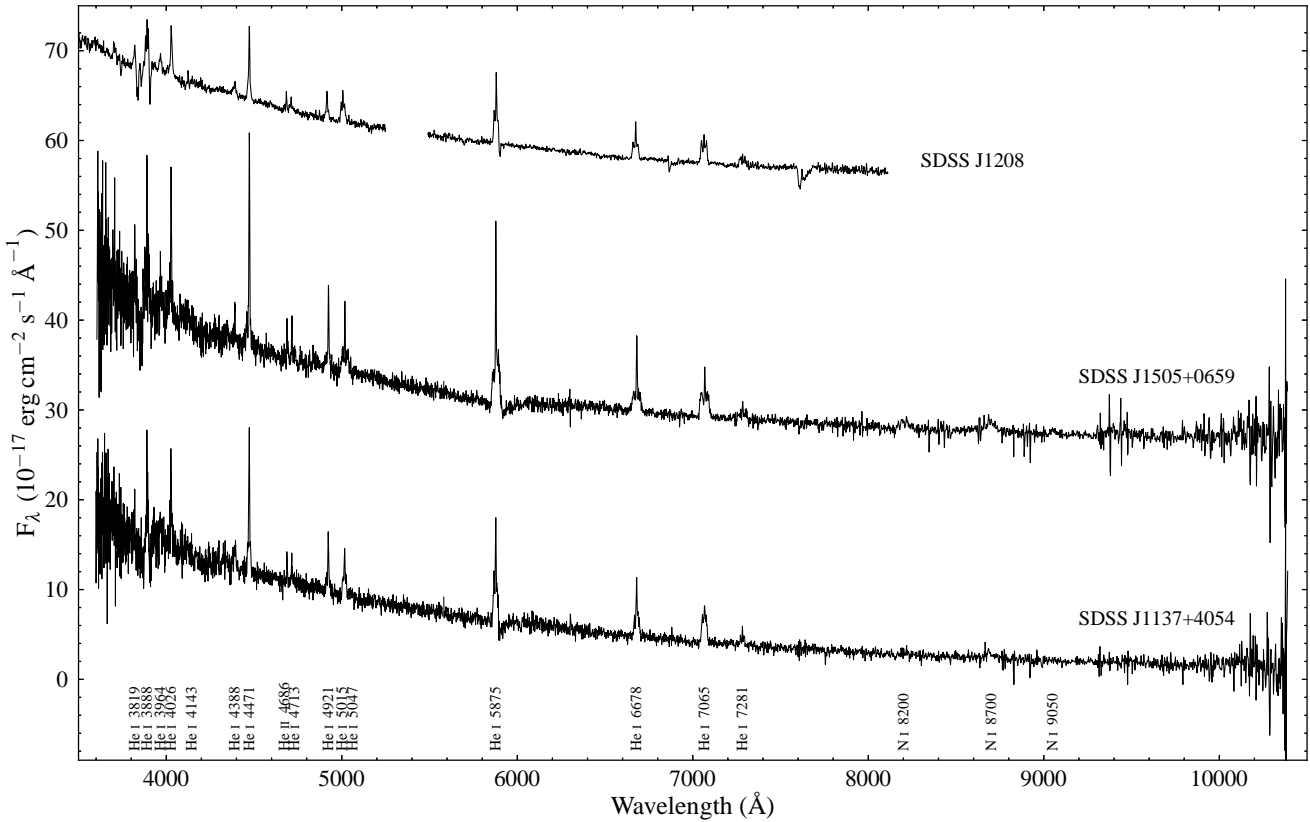}
 \caption{\textcolor{black}{SDSS spectra of the two new AM CVn binaries and SDSS\,J1208 \citep{2013MNRAS.432.2048K}. The prominent emission lines have been labelled. The spectra of SDSS\,J1505 and SDSS\,J1208 have been offset by 25\,$\times$\,10$^{-17}$ and 54\,$\times$\,10$^{-17}$\,erg\,cm$^{-2}$\,s$^{-1}$\,\AA$^{-1}$ respectively.} \label{f:spectra}}
\end{figure*}
We show the SDSS spectra of the two new AM CVn systems, SDSS\,J113732.32+405458.3 (hereafter SDSS\,J1137) and SDSS\,J150551.58+065948.7 (hereafter SDSS\,J1505) in Fig. \ref{f:spectra}. These were identified via the strong helium emission lines they display. \textcolor{black}{Their spectra are both remarkably similar to the spectrum of SDSS\,J1208 from \citet{2013MNRAS.432.2048K}, which is reproduced in Fig. \ref{f:spectra} for reference.}
Both SDSS\,J1137 and SDSS\,J1505 show triple-peaked emission lines, combining the standard double-peak associated with the accretion disc \citep{1986MNRAS.218..761H}, and a `central spike'. This feature has been seen in several AM CVn systems, and is thought to originate close to the surface of the accreting white dwarf \citep{1999MNRAS.304..443M,2003A&A...405..249M}.

\begin{table*}
\begin{minipage}{168mm}
\centering
\caption{Spectral energy distribution and equivalent width (EW) of the He\,\textsc{i} 5875\,\AA\ emission line for the two new AM CVn binaries. Numbers in parentheses indicate uncertainties in the corresponding number of last digits.}
\label{t:SED}
\begin{tabular}{l c c c c c c c c c}
\hline
Object			& FUV		& NUV		& $u$		& $g$		& $r$		& $i$		& $z$		& $A(g)$ 	& EW He\,\textsc{i} 5875 (\AA)\\
\hline
SDSS\,J113732.32+405458.3	& --	   	& 19.87(9)	& 19.26(3)	& 19.00(1)	& 19.23(1)	& 19.39(2)	& 19.45(6)	& 0.07		& $-$30(3) \\
SDSS\,J150551.58+065948.7	& 22.69(23)	& 20.05(17)	& 19.30(3)	& 19.11(1)	& 19.17(1)	& 19.21(2)	& 19.35(5)	& 0.12		& $-$50(3) \\
\hline\end{tabular}
\end{minipage}
\end{table*}
There are signs of underlying absorption around the 5875\,\AA\ emission line in both spectra. Similar features have been seen in several AM CVns \citep[e.g. SDSS\,J1240, V406 Hya][]{2006MNRAS.365.1109R}, where this is identified as being from the accretor, but it is usually clearest in the bluest lines.
Both systems also show broad nitrogen emission lines in the red parts of their spectra (see Fig. \ref{f:spectra}).

Table \ref{t:SED} gives the SDSS and \textit{GALEX} photometric magnitudes for the two systems.
\begin{figure}
 \includegraphics[width=0.45\textwidth]{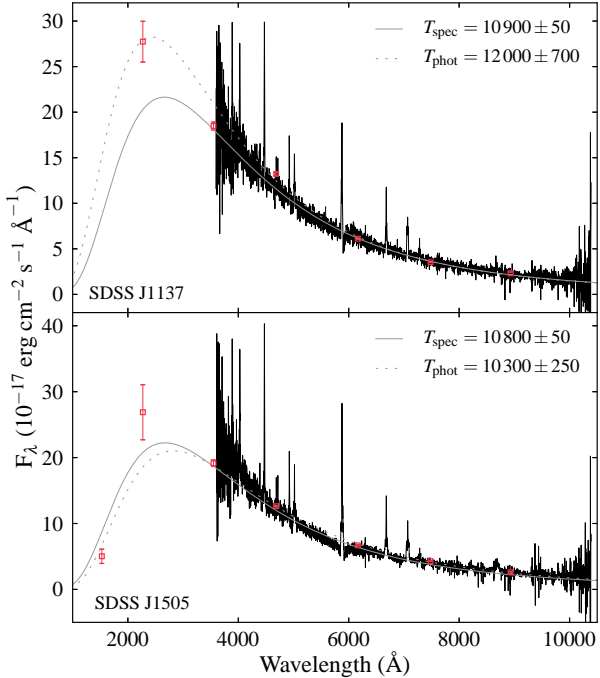}
 \caption{Spectra of the two new AM CVn binaries with photometric data from SDSS and \textit{GALEX} overplotted (red squares). The solid and dashed lines are blackbody fits to the spectra and photometry respectively. Temperatures of these fits are given in Kelvin. Data have been corrected for Galactic extinction\label{f:temperature}}
\end{figure}
We fitted a blackbody to both the spectrum (masking the emission lines) and the photometric fluxes for each object in order to estimate the continuum temperature. The data have been corrected for extinction  using the full Galactic reddening according to \citet{1998ApJ...500..525S}, $E(B-V) = 0.018$ and $0.032$ for SDSS\,J1137 and SDSS\,J1505 respectively. The fits and temperatures are shown in Fig. \ref{f:temperature}. \textcolor{black}{We give the formal uncertainties from the fits. Note that the contribution of the disc to the continuum is unknown, and is expected to vary significantly with orbital period \citep[see e.g.][]{2004MNRAS.349..181N,2006ApJ...640..466B}.}

\subsection{Radial velocities}

The SDSS spectra (Fig. \ref{f:spectra}) are the average of at least three exposures, known as sub-spectra, with typical exposure times of 15\,min. There are six sub-spectra for SDSS\,J1137, and four for SDSS\,J1505. We searched for radial velocity variations between the SDSS subspectra in order to search for the orbital periods of these systems.

We measured the radial velocities of the nine strongest He\,\textsc{i} emission lines\footnote{The following He\,\textsc{i} emission lines were used to measure the radial velocities: 3888, 4026, 4471, 4921, 5015, 5875, 6678, 7065 and 7281\,\AA} using \textsc{molly}\footnote{\textsc{molly} was written by T. R. Marsh and is available from http://www.warwick.ac.uk/go/trmarsh/software.}. Each line was fitted by a single Gaussian, with the initial values of the fit parameters determined from a fit to the average spectrum. The FWHM for each line was fixed to this value, whilst the velocity offset common to all lines, and the height of each line were allowed to vary. This gives a radial velocity for each subspectrum that is much more precise than achievable using any single line.

\textcolor{black}{The measured radial velocities, given in Tables \ref{t:rvs1137} and \ref{t:rvs1505}, were then used to construct power spectra for each object, to search for periodic signals. We use both the Lomb-Scargle \citep{1976Ap&SS..39..447L,1982ApJ...263..835S} and orthogonal polynomial \citep[ORT;][]{1996ApJ...460L.107S} methods to calculate power spectra.}
\begin{table}
\centering
\caption{\textcolor{black}{Radial velocities measured from the SDSS\,J1137 subspectra.}}
\label{t:rvs1137}
\begin{tabular}{l r}
\hline
HJD		& Radial velocity (km s$^{-1}$) \\
\hline
2455620.8272816		& 38.20 $\pm$ 16.85 \\
2455620.8386465		& 102.47 $\pm$ 19.84 \\
2455620.8500112		& -24.08 $\pm$ 18.58 \\
2455620.8613760		& -13.26 $\pm$ 19.39 \\
2455620.8727523		& 95.37 $\pm$ 14.75 \\
2455620.8841171		& 89.73 $\pm$ 14.71 \\
\hline
\end{tabular}
\end{table}
\begin{table}
\centering
\caption{\textcolor{black}{Radial velocities measured from the SDSS\,J1505 subspectra.}}
\label{t:rvs1505}
\begin{tabular}{l r}
\hline
HJD		& Radial velocity (km s$^{-1}$) \\
\hline
2455712.6830284		& 76.53 $\pm$ 27.44 \\
2455712.6944161		& 54.92 $\pm$ 28.41 \\
2455712.7057808		& 87.89 $\pm$ 26.39 \\
2455712.7171455		& 94.27 $\pm$ 24.99 \\
\hline
\end{tabular}
\end{table}
\begin{figure}
 \includegraphics[height=0.45\textwidth,angle=270]{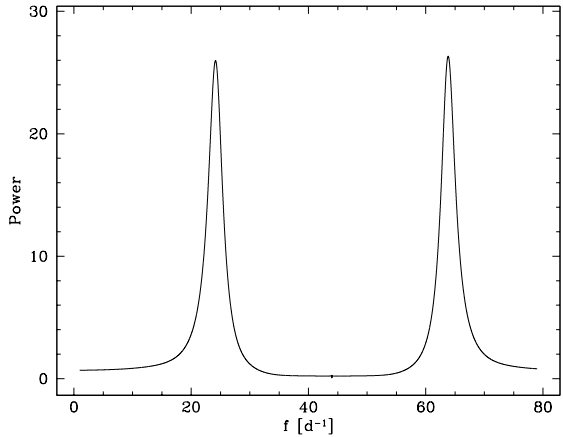}
 \caption{\textcolor{black}{ORT power spectrum calculated from the He\,\textsc{i} radial velocities of SDSS\,J1137. Two strong peaks of almost equal power are seen at 24.1 and 63.8\,cycles\,d$^{-1}$.}\label{f:powerspec}}
\end{figure}
\begin{figure}
 \includegraphics[height=0.45\textwidth,angle=270]{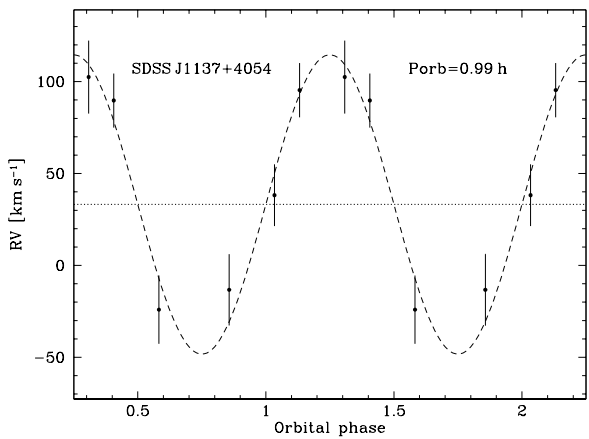}
 \caption{Measured He\,\textsc{i} radial velocities for SDSS\,J1137 folded on a period of 59.6\,minutes. The dashed and dotted lines are the best fitting radial velocity curve and systemic velocity.\label{f:RV1137}}
\end{figure}
\textcolor{black}{The power spectrum for SDSS\,J1137 (shown in Fig. \ref{f:powerspec}) is dominated by two strong peaks, corresponding to periods of 22.56\,$\pm$\,0.39 and 59.63\,$\pm$\,2.74\,min}. The phase folded radial velocity curve for the longer period is shown in Fig. \ref{f:RV1137}.

The power spectra for SDSS\,J1505 produce a similar result, with the stronger peak corresponding to a period of 50.6\,min. However, with only four data points, the low radial velocity amplitude and the large uncertainties on the measurements, we do not consider this to be a reliable measurement of the orbital period.


\section{Discussion}

The spectra of the two new AM CVn binaries are very similar to that of the long period AM CVn SDSS\,J1208 \citep[$P_{\rmn{orb}}\,=\,53$\,min;][]{2008AJ....135.2108A,2013MNRAS.432.2048K} showing broad double-peaked helium emission with strong central spikes (Fig. \ref{f:spectra}). Due to the mass transfer rate being a steep function of the orbital period, the spectra of AM CVn binaries vary considerably with period. That these new systems have spectra that appear similar to SDSS\,J1208 suggests that they may also be long period AM CVn binaries ($\gtrsim$50\,min).

\textcolor{black}{The two possible periods determined from radial velocities of the subspectra for SDSS\,J1137 have almost equal power in the power spectra. At an orbital period of 22.5\,min we would expect the system to spend much of its time in a high state, or to undergo frequent outbursts (similarly to KL Dra; \citealt{2012MNRAS.419.2836R}). On the other hand, at a period of 59.6\,min we would not expect to see any outbursts, with the disc being in a stable quiescent state. As there is no evidence of outbursts (or other significant variability) in the 9 year coverage of the Catalina Real Time Transient Survey (CRTS; \citealt{2009ApJ...696..870D}) lightcurve, and the similarity of SDSS\,J1137's spectrum to that of the long period system SDSS\,J1208, gives support to the longer of the two possible periods.}

\textcolor{black}{The CRTS lightcurve for SDSS\,J1505 also shows no sign of any outbursts (though infrequent outbursts could have been missed), and the spectrum is also very similar to that of SDSS\,J1208 (Fig. \ref{f:spectra}), again suggesting a long period for this system.
We do not consider the periodic signal measured for SDSS\,J1505 to be a reliable measure of the orbital period due to the paucity of data, but it might lend further support to the suggestion that it is a another long period system.}

\textcolor{black}{The chemical abundances of material in the accretion discs of AM CVns can, in principle, be used to infer the formation channel of these systems (\citealt{2010MNRAS.401.1347N}; defined by the evolutionary state of the donor at the onset of Roche-lobe overflow, see e.g. \citealt{2010PASP..122.1133S}).} The detection of nitrogen emission in the spectra of AM CVn binaries has been used to suggest that they are more likely to have white dwarf donors than helium star donors \citep[e.g.][]{2013MNRAS.430..996L,2014MNRAS.437.2894C}. We cannot constrain the N/C ratio as there are no strong carbon lines in the optical, but the detection of N\,\textsc{i} lines likely rules out significantly evolved He star donors \citep{2010MNRAS.401.1347N}.

Both spectra show signs of absorption wings around the He\,\textsc{i} 5875 emission line (especially SDSS\,J1505), that could be associated with the accreting white dwarf. However, this would require a white dwarf temperature of $\sim$15\,000\,K or higher, which does not seem to be compatible with the NUV flux detected for either system. It is also worth noting that at a period longer than 50\,min, such a temperature would be unexpectedly high, as the white dwarf should cool as the orbital period increases \citep{2006ApJ...640..466B}. We also note that where DB absorption wings have been observed in other AM CVn binaries, it has been strongest around bluer lines, particularly He\,\textsc{i} 4471, and has not been noticeable around the 5875\,\AA\ line \citep[e.g.][]{2006MNRAS.365.1109R}. This feature may be due to interstellar Na absorption, SDSS\,J1137 shows only a red wing, and the red wing appears stronger in SDSS\,J1505.

The continuum temperatures suggest cool accretors for both systems, this would be expected for long period systems, and helps to explain the redder than average $u-g$ colours that push these objects outside of our original colour box. Even considering the low temperature of the accretor, SDSS\,J1137 has an extremely red $u-g$ colour that puts it surprisingly close to the DA white dwarf track in $u-g$, $g-r$ colour space (see Fig. \ref{f:newcuts}).

That we found no further new systems in our search of the SDSS spectroscopic data base leads us to suspect that these two systems are outliers from the main colour distribution of AM CVn binaries. If they represent an additional population that is missed in our earlier systematic search \citep{2013MNRAS.429.2143C}, then this population is likely small, since there are only two such systems in the SDSS spectroscopic data base (compared to the nine systems with SDSS spectra in \citealt{2009MNRAS.394..367R}'s original colour box). The discovery of these two new AM CVns should have very little effect upon previous estimates of the AM CVn space density.


\section*{Acknowledgements}

We thank the anonymous referee for helpful comments and suggestions. PJC and NPGF acknowledge the support of Science and Technology Facilities Council (STFC) studentships. The research leading to these results has received funding from the
European Research Council under the European Union's Seventh Framework
Programme (FP/2007-2013) / ERC Grant Agreement n. 320964 (WDTracer).
BTG was supported in part by the UK's STFC (ST/I001719/1).
DS, TRM and EB acknowledge support from the STFC grant no. ST/I001719/1. TK acknowledges support by the Netherlands Research School for Astronomy (NOVA). GN acknowledges an NWO-VIDI grant.
Funding for SDSS-III has been provided by the Alfred P. Sloan Foundation, the Participating Institutions, the National Science Foundation, and the U.S. Department of Energy Office of Science. The SDSS-III web site is http://www.sdss3.org/.
This research has made use of NASA's Astrophysics Data System.
Fig. \ref{f:newcuts} makes use of P. Bergeron's synthetic white dwarf colours \citep{2006AJ....132.1221H,2006ApJ...651L.137K,2011ApJ...730..128T,2011ApJ...737...28B}, kindly made available by the authors at http://www.astro.umontreal.ca/$\sim$bergeron/CoolingModels.


\label{lastpage}

\end{document}